\documentclass[twocolumn,superscriptaddress,showpacs,oneside,amsmath,amssymb,prl]{revtex4}
\usepackage{epsfig,amssymb,amsfonts}

\begin{document}

\title{Spin Fluctuations, Interband Coupling, and Unconventional Pairing in
  Iron-based Superconductors}

\author{Zi-Jian Yao}

\affiliation{Department of Physics and Center of Theoretical and
  Computational Physics, The University of Hong Kong, Pokfulam Road,
  Hong Kong, China}
\affiliation{National Laboratory of Solid State Microstructures
and
  Department of Physics, Nanjing University, Nanjing 210093, China}
\author{Jian-Xin Li}
\affiliation{National Laboratory of Solid State Microstructures
and
  Department of Physics, Nanjing University, Nanjing 210093, China}
\affiliation{Department of Physics and Center of Theoretical and
  Computational Physics, The University of Hong Kong, Pokfulam Road,
  Hong Kong, China}
\author{Z. D. Wang}
\affiliation{Department of Physics and Center of Theoretical and
  Computational Physics, The University of Hong Kong, Pokfulam Road,
  Hong Kong, China}

\date{\today}
\begin{abstract}
Based on an effective two-band model and using the
fluctuation-exchange (FLEX) approach, we explore spin fluctuations
and unconventional superconducting pairing in Fe-based layer
superconductors. It is elaborated that one type of interband antiferromagnetic (AF) spin fluctuation stems from the interband Coulomb repulsion, while the other type of intraband AF spin fluctuation originates from the intraband Coulomb repulsion. Due to the Fermi-surface topology, a spin-singlet extended $s$-wave superconducting state is more favorable than the nodal $d_{XY}$-wave state if the interband AF spin fluctuation is more significant than the intraband one, otherwise vice versa. It is also revealed that the effective interband coupling
plays an important role in the intraband pairings, which is a
distinct feature of the present two-band system.

\end{abstract}


\maketitle

The recent discovery of superconductivity with higher transition
temperatures in the family of iron-based materials~\cite{kam1} has
stimulated enormous research interests both
experimentally~\cite{wen,kam2,wen2,h1,h2,h3,h4,sdw1,neutron1,neutron2}
and
theoretically~\cite{sing,xu,mazin,kuroki,cao,dai,han,Eremin,raghu,wxg,weng,ma}.
In particular, the origin and nature of superconductivity and spin
density wave (SDW) ordering observed in these materials have been
paid considerable
attention~\cite{wen,wen2,sdw1,neutron1,neutron2,sing,mazin,kuroki,dai,han,weng,yin}.
Currently available experimental data suggested that the
superconducting pairing state exhibit nodal behaviors~\cite{wen2},
while preliminary theoretical arguements/analyses indicated the
pairing possibilities of an extended $s$-wave (either
without~\cite{mazin} or with nodes~\cite{kuroki}), a nodal
$d$-wave~\cite{kuroki,han}, a spin-triplet $s$-wave~\cite{dai}, and
a spin-triplet $p$-wave~\cite{wxg}, all of them are based on the
scenario that spin fluctuations induce the superconductivity in this
kind of systems (with antiferromagnetic fluctuations being
responsible for the former two spin-singlet pairings while
ferromagnetic origin for
the latter two spin-triplet pairings). 
Therefore, systematic and profound theoretical investigations on
spin fluctuations and their relationship with the superconducting
pairing are significant and of current interest.

This new family of superconductors has a layered structure, where
the FeAs layer is experimentally suggested  to be responsible for
the superconductivity~\cite{kam1,wen,kam2,h1,h2,h3,h4}. The LDA band
calculations~\cite{sing,xu,yin} indicate that there are five bands
intersect the Fermi level in the folded Brillouin zone (BZ), in
which four bands are quasi-two-dimensional. Therefore, in the
representation  of the unfolded (or extended) BZ, two bands may be
able to reproduce the main features of the four Fermi pockets after
folding. In this paper, we employ an effective
 two-band model Hamiltonian~\cite{raghu} to explore
 the low energy excitation physics
 including spin fluctuations and superconducting pairing with
 the FLEX approach~\cite{bickers}. It is illustrated that one type of
 commensurate AF spin fluctuations stems from the interband
Coulomb repulsion associated with the nesting between the electron
and hole Fermi pockets, while the other type of intraband AF spin
fluctuation originates from the intraband Coulomb repulsion. Due to
the Fermi-surface topology, a spin-singlet extended $s$-wave
superconducting state is more favorable than the nodal $d_{XY}$-wave
state as the interband spin fluctuation is significant, otherwise
vice versa. It is also elaborated that the effective interband
coupling is enhanced by the interband AF spin fluctuation and plays
an important role in the intraband pairings.


 We start from an effective two-band model Hamiltonian
\begin{equation}
  H=H_0+H_{int},
\end{equation}
where $H_0$ is given by
\begin{equation}
H_{0}=\sum_{kl\sigma}\varepsilon_{l}(k)c^{\dagger}_{kl\sigma}c_{kl\sigma}.
\end{equation}
Here $c_{kl\sigma}$ denotes the band-electron annihilation operator
with the wave vector $k$, spin $\sigma$ in the band $l$ ($l=1,2$).
In the present work, the two energy bands denote the hole band
(band-$1$) and electron band (band-$2$), with their dispersions
being approximated by $\varepsilon_{1,2}(k)=[\xi_{xz}+\xi_{yz}\mp
\sqrt {(\xi_{xz}-\xi_{yz})^{2}+4\epsilon^{2}}]/2$,
where $\xi_{xz}(k)=-2t_1\cos k _x-2t_2\cos k_y-4t_3\cos k_x\cos
k_y,$ $\xi_{yz}(k)=-2t_2\cos k_x-2t_1\cos k_y-4t_3\cos k_x\cos k_y,$
$\epsilon(k)=-4t_4\sin k_x\sin k_y$, as addressed in
Refs.~\cite{raghu,graser}. Here, $\xi_{xz}(k)$ and $\xi_{yz}(k)$ may
be understood as the iron $d_{xz}$ and $d_{yz}$ orbit-dispersions
while $\epsilon(k)$ as the hybridization of the two orbits.
\begin{figure}
\includegraphics[scale=0.5]{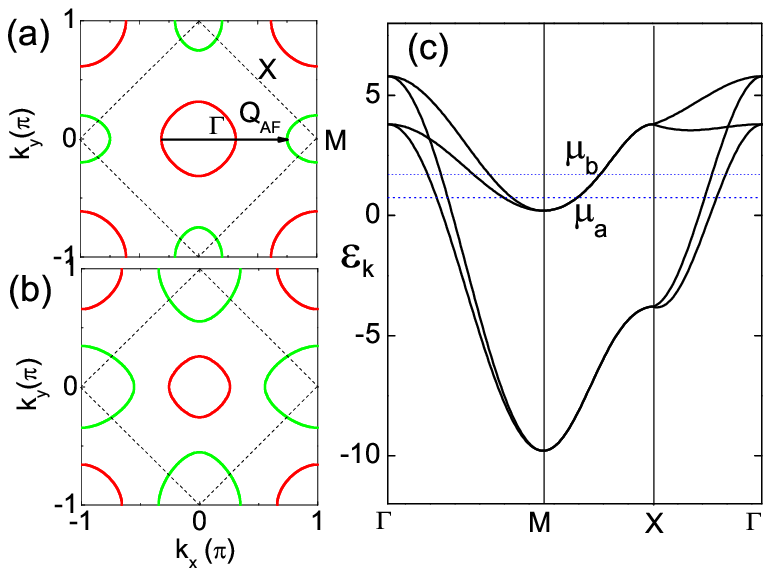}
\caption{(Color online) (a) and (b): The Fermi surfaces of the
two-band model in the extended Brillouin zone (1 Fe per cell) for
$\mu_a=0.74$ and $\mu_b=1.7$, where the thin dashed line denotes the
folded Brillouin zone (2 Fe per unit cell) and the arrow represents
the nesting wave vector (see text). (c) The corresponding band
structure (energy is in the unit of $t_1$) in the folded Brillouin zone.}
\end{figure}
To produce better the topology of the Fermi surface and band
features of the LDA  calculations~\cite{sing,xu,yin}, we set
$t_1=-1.0,\, t_2=1.5,\, t_3=-1.2,\, t_4=-0.95, \,\mu=0.74,\,1.7$
(in units of $|t_1|$), which gives rise to the electron Fermi
pockets and the hole Fermi pockets (being respectively denoted by
the green and red lines in Fig.1 (a)) as well as the band
structure (Fig.1(c)).

The interacting term $H_{int}$ consists of the effective intraband
Coulomb interaction~\cite{note},
$(U/2)\sum_{i,l,\sigma\neq\sigma^{\prime}}c^{\dagger}_{il\sigma}
c^{\dagger}_{il\sigma^{\prime}}c_{il\sigma^{\prime}}c_{il\sigma}$,
the effective interband Coulomb interaction
$(U^{\prime}/2)\sum_{i,l\neq
l^{\prime},\sigma,\sigma^{\prime}}c^{\dagger}_{il\sigma}
c^{\dagger}_{il^{\prime}\sigma^{\prime}}c_{il^{\prime}\sigma^{\prime}}c_{il\sigma}$,
the Hund's coupling $J\sum_{i,l\neq
l^{\prime},\sigma\sigma^{\prime}}c^{\dagger}_{il\sigma}
c^{\dagger}_{il^{\prime}\sigma^{\prime}}c_{il\sigma^{\prime}}c_{il^{\prime}\sigma}$
, and the interband pair-hopping term $J^{\prime}\sum_{i,l\neq
l^{\prime},\sigma\neq\sigma^{\prime}}c^{\dagger}_{il\sigma}
c^{\dagger}_{il\sigma^{\prime}}c_{il^{\prime}\sigma^{\prime}}c_{il^{\prime}\sigma}$,
where the $i$-site is defined on the reduced lattice (one Fe per
cell).


Experimental data~\cite{kam2,sdw1,neutron1,neutron2} indicated
that the undoped material LaOFeAs behaves like a semimetal, and
exhibits the itinerant antiferromagnetism. Thus it is reasonable
to consider the Coulomb interaction to be intermediate in this
system. In this sense, the FLEX approach~\cite{bickers} appears to
be an adequate method. In this approach, the spin/charge
fluctuations and the electron spectra are determined
self-consistently by solving the Dyson's equation with a primary
bubble- and ladder-type effective interaction. For the two-band
system, 
the Green's function and the self-energy are expressed as the
$2\times2$ matrices, satisfying the Dyson equation:
$\hat{G(k)}=[i\omega_n\hat{I}-\hat{\varepsilon}(k)
-\hat{\Sigma}(k)]^{-1}$, with $\varepsilon_{11}=\varepsilon_{1}$,
$\varepsilon_{22}=\varepsilon_{2}$ and
$\varepsilon_{12}=\varepsilon_{21}=0$. The self-energy reads
$\Sigma_{mn}(k)=\frac{T}{N}\sum_{q}\sum_{\mu\nu}V_{{\mu}m,{\nu}n}(q)G_{\mu\nu}(k-q)$,
where the effective interaction $\hat{V}$ is a $4\times4$ matrix (with
the basis ($\left|11\right\rangle$, $\left|22\right\rangle$,
$\left|12\right\rangle$, $\left|21\right\rangle$)) given
by~\cite{takimoto}
\begin{eqnarray}
V_{{\mu}m,{\nu}n}(q)&=&[\frac{3}{2}\hat{U}^s\hat{\chi}^s(q)\hat{U}^s+\frac{1}{2}\hat{U}^c\hat{\chi}^c(q)\hat{U}^c+\frac{3}{2}\hat{U}^s\\
\nonumber
&-&\frac{1}{2}\hat{U}^c-\frac{1}{4}(\hat{U}^s+\hat{U}^c)\hat{\overline{\chi}}(\hat{U}^s+\hat{U}^c)]_{{\mu}m,{\nu}n},
\end{eqnarray}

with
\begin{equation}
\hat{\chi}^s({q})=[\hat{I}-\hat{\overline{\chi}}(q)\hat{U}^s]^{-1}\hat{\overline{\chi}}(q),\,
\hat{\chi}^c({q})=[\hat{I}+\hat{\overline{\chi}}(q)\hat{U}^c]^{-1}\hat{\overline{\chi}}(q)
\end{equation}
as the spin and charge fluctuations. The irreducible
susceptibility is
$\overline{\chi}_{{\mu}m,{\nu}n}(q)=-\frac{T}{N}\sum_{k}G_{{\nu}{\mu}}(k+q)G_{mn}(k)$,
and the interaction vertex reads,
\begin{equation}
\hat{U}^{s}=\left(\begin{array}{cc} \hat{U}^{s1} & 0\\
0 & \hat{U}^{s2}\end{array}\right),
\hat{U}^{c}=\left(\begin{array}{cc} \hat{U}^{c1} & 0\\
0 & \hat{U}^{c2} \end{array}\right), \label{tew}
\end{equation}
where $\hat{U}^{s1}_{mn}=U$ for $m=n$ and $2J$ otherwise,
$\hat{U}^{s2}_{mn}=U^{\prime}$ for $m=n$ and $2J^{\prime}$
otherwise, $\hat{U}^{c1}_{mn}=U$ for $m=n$ and $2U^{\prime}-2J$
otherwise, $\hat{U}^{c2}_{mn}=-U^{\prime}+4J$ for $m=n$ and
$2J^{\prime}$ otherwise. In the above equations,
$k\equiv(\mathbf{k}, \mathbf{\omega}_n)$ and $q\equiv(\mathbf{q},
i\mathbf{\nu}_n)$ are used, with $\omega_{m}$ the Matsubara
frequency, $T$ the temperature, and $N$ the lattice site number.

The above equations form a closed set of equations and can be solved
self-consistently to get the renormalized Green's function in the
presence of the interaction $H_{int}$. After obtaining $\hat{G}(k)$,
we can look into the superconducting instability and the gap
symmetry from the following Eliashberg equation~\cite{takimoto}
\begin{eqnarray}
  \lambda\Delta_{mn}(k)&=&-\frac{T}{N}\sum_q\sum_{\alpha\beta}\sum_{\mu\nu}\Gamma^{s,t}_{{\alpha}m,n\beta}(q)
  \\ \nonumber
  & &\times G_{\alpha\mu}(k-q)G_{\beta\nu}(q-k)\Delta_{\mu\nu}(k-q)
\end{eqnarray}
with the pairing potential being given by $
\hat{\Gamma}^s(q)=\frac{3}{2}\hat{U}^s\hat{\chi}^s(q)\hat{U}^s-\frac{1}{2}\hat{U}^c\hat{\chi}^c(q)\hat{U}^c+\frac{1}{2}(\hat{U^s}+\hat{U}^c)
$ and
$\hat{\Gamma}^t(q)=-\frac{1}{2}\hat{U}^s\hat{\chi}^s(q)\hat{U}^s-\frac{1}{2}\hat{U}^c\hat{\chi}^c(q)\hat{U}^c+\frac{1}{2}(\hat{U^s}+\hat{U}^c)
$ for the spin-singlet and spin-triplet states ($\hat{\Gamma}^t(q)$
is the same for the pairing spin projection $S_z=\pm 1,0$), respectively. The
eigenvalue $\lambda\rightarrow 1$ when the supercondcuting
transition temperature $T_c$ is reached. It is worth indicating that
the interband Cooper pairing gap function $\Delta_{12}$ is decoupled
from the equation of intraband Cooper pairing gap functions
$\Delta_{ll}$ ($l=1,\,2$) and is vanishingly small for the present
Fermi pockets pattern.

Numerical calculations are carried out with $32\times 32$ $k$-meshes
in the extended BZ and 1024 Matsubara frequencies. The analytic
continuation to the real frequency is carried out with the usual
Pad$\acute{e}$ approximant. As for the interaction parameters, we
note that a set of parameters with $U=0.2-0.5$ bandwidth and
$J\approx 0.09$ bandwidth was used in the literature~\cite{cao}.
Here we choose $8$ sets of representative parameters: ($U,\,
U^{\prime}$)=(6.5, 3.5) and (5.5, 4.0) for $J^{\prime}=1.0\, \& 0.5$
and $\mu=0.74\,\&1.7$ with $J=1.0$.

Let us first address the static spin susceptibility $\chi^{s}
(\omega=0)$ ($\chi^{c}(\omega=0)\ll\chi^{s} (\omega=0)$, not shown
here.). Fig.2 presents the physical spin susceptibility
$\chi^{s}_{ph}=\sum_{mn}\chi^{s}_{mn,mn}$, its intraband
components $\chi^s_{22}$, $\chi^s_{11}$, and the interband one
$\chi^s_{12}$ in the extended BZ with $U=6.5, \, U^{\prime}=3.5,
\, J=J^{\prime}=1.0$, and $\mu=0.74$. (For brevity,
$\chi^{s}_{mn}\equiv\chi^{s}_{mn,mn}$ is used hereafter.) The
physical spin susceptibility displays two sets of dominant peaks,
with one around $(\pi,0)$ and its symmetric points in the extended
BZ, and the other around $(0.5\pi,0.5\pi)$ and its symmetric
points. The commensurate AF spin fluctuation around $(\pi,0)$
comes from the interband Coulomb interaction associated with the
nesting between the hole and electron pockets~\cite{han} (Fig.1
(a)), which is clearly seen from the interband component
$\chi^{s}_{12}$ shown in Fig.2(d). The appearance of this type of
AF spin fluctuation around $(\pm\pi,0)$ and $(0,\pm\pi)$ is in
good agreement with the neutron scattering
measurements~\cite{neutron1,neutron2}, and here we refer to it as
the interband AF spin fluctuation. More intriguingly, the other
type of intraband spin fluctuation is seen to peak around
$(0.5\pi,0.5\pi)$ in the components $\chi^s_{11}$ and
$\chi^s_{22}$, which is mainly induced by the intraband Coulomb
interaction $U$ associated with the approximate nesting property
within the renormalized Fermi pocket around the $(\pi,\pi)$ point
(not shown here).
Notably, the peak position corresponds to $[0, \pi]$ (and $[\pi,
0]$) in the folded BZ and thus implies the emergence of a new
component of "stripe"-type AF spin fluctuation in the primary
lattice (with 2 Fe ions per unit cell), which could be referred to
as the intraband spin fluctuation and may be detected directly by
future neutron scattering experiments on single crystal samples.

\begin{figure}
   \includegraphics[scale=0.5]{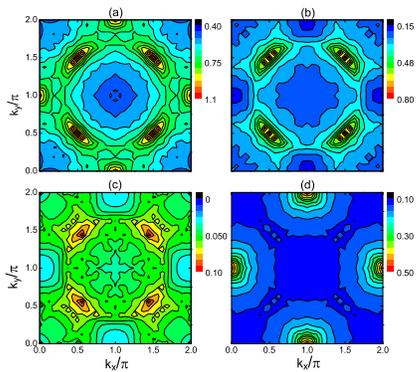}
   \caption{(Color online) The $k$-dependence of the static spin susceptibility
   for
   $U=6.5,U^{\prime}=3.5,\,J=J^{\prime}=1$, and $\mu=0.74$ at temperature $T=0.01$.
   (a) The physical spin susceptibility (see text). (b)-(d) The components of the spin susceptibility
   $\chi^{s}_{22}$, $\chi^{s}_{11}$, and $\chi^{s}_{12}$, respectively.
   }
\end{figure}

\begin{figure}
   \includegraphics[scale=0.7]{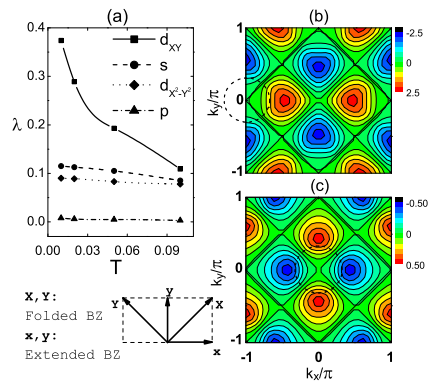}
    \caption{(Color online) (a): Temperature dependence of the maximum
    eigenvalues,   
    (b) and
    (c): $k$-dependence of the
     gap functions $\Delta_{2,1}(k)$ corresponding to the largest
     eigenvalue, for the same set of parameters as those in Fig.2 at temperature $T=0.01$.
     The solid diamond is the folded
     BZ and
     the dashed circle denotes the Fermi pocket schematically.}
\end{figure}
The most favorable superconducting pairing symmetry at a fixed
temperature is determined by solving the Eliashberg equation with
the maximum eigenvalue. The calculated  maximum eigenvalues (for
various possible pairing symmetries) versus temperature are plotted
in Fig.3(a). Firstly, one can see that the eigenvalue for the
spin-triplet $p$-wave state is much smaller than those of the
spin-singlet state, and at the mean time exhibits a flat temperature
dependence. Therefore, we can safely rule out the possibility of the
spin-triplet state in the present model calculation. In the
spin-singlet channel, the eigenvalue of $d_{XY}$-wave state is
larger than that of the $s$-wave state,
and in particular the former increases rather rapidly with
decreasing temperature. In view of this tendency, although the
maximum eigenvalue $\lambda=1$ has not been reached yet, it is
reasonable to consider the spin-singlet $d_{XY}$-wave to be the most
favorable state in this set of parameters. The calculated $k$-space
structure of the gap functions for both the electron and hole bands
in the extended BZ are depicted in Figs.3(b) and (c), respectively.
It is seen that the pairing symmetries in both bands are of
$d_{XY}$-wave, namely, the gap function
$\Delta_{ll}(\mathbf{k})\approx \Delta^0_{l} \gamma_{l\mathbf{k}}$
with $\gamma_{l\mathbf{k}} \approx 2 \sin k_X \sin k_Y$, where
($k_X$, $k_Y$) is the wave vector denoted in the folded BZ.
Interestingly, we note that the gap magnitude in the electron-band
is significantly larger than that in the hole-band.
This feature could also be understood as follows. Eq.(6) may
approximately  be rewritten as
\begin{equation}
\lambda \Delta^0_{l}=\sum_{l^{\prime}} K_{ll^{\prime}}
\Delta^0_{l^{\prime}}
\end{equation}
for $\omega=0$, where
$K_{ll^{\prime}}=\sum_{\mathbf{k}, \mathbf{k}^{\prime}}
\tilde{V}_{ll^{\prime}}(\mathbf{k}-\mathbf{k}^{\prime}) $
with the effective intraband pairing potential and interband
coupling as
$\tilde{V}_{ll^{\prime}}=-|G_{l^{\prime}l^{\prime}}|^2[(U^{2}+4J^{2})\chi^s_{ll^{\prime}}\delta_{ll^{\prime}}+
4U^{\prime}J^{\prime}\chi^s_{ll^{\prime}}(1-\delta_{ll^{\prime}})]\gamma_{l\mathbf{k}}
\gamma_{l^{\prime}\mathbf{k}^{\prime}}/(N\sum_{\mathbf{k}}
\gamma^2_{l\mathbf{k}})$. Since $K_{22}\gg K_{11}$ (mainly due to
the result $\chi^s_{22} \gg \chi^s_{11}$), the gap amplitude of
electron band (band 2) is mainly determined from the intraband
pairing of itself, while that of the hole-band depends mainly on the
interband coupling from the electron band. From the expression of
$\tilde{V}_{ll^{\prime}}$, it is elucidated that the intraband spin
AF fluctuation leads to the intraband pairings, while the interband
AF spin fluctuation enhances the effective interband coupling
(associated with the interband coupling factor
$U^{\prime}J^{\prime}$) and thus the intraband pairings of both
bands~\cite{note2}. The above conclusions are unchanged for
$\mu=1.7$ or $J^{\prime}=0.5$, where the intraband spin fluctuation
has an even more impact on the pairing. However, if the interband AF
spin fluctuation is dominant over the intraband one,
 as seen in Fig.4(a)
for $U=5.5,\,U^{\prime}=4.0$, $J=J^{\prime}=1$, and $\mu=0.74$,
an extended $s$-wave state (Figs.4 (b) and (c)) would be more
favorable than the $d$-wave one. A similar conclusion has also be
obtained in a recent renormaliztion group study~\cite{fawang}.


Because the spin fluctuation $\chi^{s}$ is stronger than the charge
fluctuation $\chi^{c}$ (not shown here), the pairing interaction in
the spin-singlet channel is positive ($\hat{\Gamma}^s(q)>0$).
Consequently, once the gap function satisfies the condition
$\Delta_l(\mathbf{k})\Delta_{l^\prime}(\mathbf{k}+\mathbf{Q})<0$,
where ${\bf Q}$ is the wave-vector around which the pairing
interaction peaks,
we are able to obtain the largest eigenvalue solution of the
Eliashberg equation.
Focusing merely on the interband AF spin fluctuation,
the above condition leads to two candidates of pairing symmetry on
the singlet channel.
    One is the
extended $s$-wave, with the gap function of each Fermi pocket having
the same sign, while changing the sign between the electron and hole
pockets, as shown in Fig.4(b) and (c). The other is the $d$-wave as
shown in Fig.3(b) and (c). Actually, which one is more favored
depends mainly on the existence of the intraband spin fluctuation
caused by the Coulomb interaction $U$.
For a larger $U$ and the approximate nesting within the
renormalized Fermi pocket, this spin fluctuation that peaks at
$(0.5\pi,0.5\pi)$ emerges (see Fig.2(b) and (c)), which leads to
the gap function to change its sign within each Fermi pocket and
thus induces a nodal $d_{XY}$-wave pairing. In contrast, if the
intraband Coulomb interaction $U$ is relatively weak, such that
the intraband spin fluctuation is not significant, the extended
$s$-wave state would energetically be favored as it opens a full
gap around the Fermi pockets~\cite{emerin1}. We think that this
kind of connection between the peak structure of the spin response
and the pairing symmetry
established here is useful for probing the superconducting pairing
symmetry by measuring the $k$-dependence of spin fluctuations in
neutron scattering.

Finally, we note that the LDA energy bands in this system exhibits a
more complex structure~\cite{sing,mazin,kuroki}. In the present
two-band model, the $d_{xz}$ and $d_{yz}$ orbitals are chosen as
these two orbitals have the largest weights to the energy bands
crossing the Fermi level~\cite{boeri}. While, the $d_{xy}$ orbital
 contributes also a weight to the energy band along the $\Gamma-M$
direction~\cite{boeri}, and particularly in the unfolded Brillouin
zone one of the hole Fermi pockets is displaced from the $(0,0)$
point to the $(\pi,\pi)$ point as depicted in Figs.1(a) and (b),
which is actually different from that in a more realistic energy
band structure~\cite{kuroki}. However, we wish to indicate that
these two deficiencies do not affect meaningfully the nesting
properties of the Fermi pockets, and thus the results/conclusions
obtained above are still unchanged, at least qualitatively.

\begin{figure}
      \includegraphics[scale=0.8]{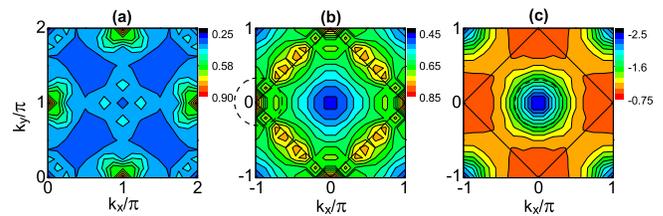}
    \caption{(Color online) (a): The physical spin susceptibility, (b) and (c): $k$-dependence of the
     gap functions $\Delta_{2,1}(k)$ corresponding to the largest
     eigenvalue, for $U=5.5,U^{\prime}=4.0,\,J=J^{\prime}=1.0$, and $\mu=0.74$ at temperature $T=0.01$.}

\end{figure}
In summary, based on an effective two-band model and using the
fluctuation-exchange approach, we have investigated spin
fluctuations and superconductivity as well as the interband coupling
in iron-based layered superconductors. We have elaborated that one
type of commensurate AF spin fluctuation comes from the interband
Coulomb interaction associated with the nesting between the hole and
electron Fermi pockets, while the other type of intraband AF spin
fluctuation originates from the intraband Coulomb repulsion. We have
elucidated that, 
if the interband AF spin fluctuation plays a
significant role, this fluctuation leads to the pairing with the spin-singlet extended $s$-wave being the most favorable state. Otherwise,
the pairing is mainly determined by the intraband AF spin fluctuation, with $d_{XY}$-wave symmetry.

{\it Note added}--(i) Recently, we note that the angle-resolved
photoemission spectroscopy measurement on the (Ba,K)Fe$_2$As$_2$
superconductor~\cite{ding} indicated the presence of nodeless gaps,
which might imply the interband spin fluctuation plays likely a
dominant role at least in these iron-based superconductors. (ii)
After the work was posted on the e-Print archive, we note a similar
work done by X.-L. Qi {\it et al} in ~\cite{qi}.


We thank F.C. Zhang, Q. Han, Y. Chen, X. Dai, Z. Fang, T. K. Ng, Q. H.
Wang, H. H. Wen, and S. C. Zhang for many helpful discussions. The
work was supported by the NSFC (10525415 and 10429401), the RGC grants
of Hong Kong (HKU-3/05C),
the 973 project (2006CB601002,2006CB921800), and the Ministry of
Education of China (Grants No.NCET-04-0453).


\begin{thebibliography}{40}

\bibitem{kam1} Y. Kamihara, T. Watanabe, M. Hirano, and H. Hosono,
J. Am. Chem. Soc. {\bf 130}, 3296 (2008).

\bibitem{wen} H. H. Wen, G. Mu, L. Fang, H. Yang, and X. Zhu, EPL {\bf 82}, 17009 (2008).

\bibitem{kam2} Y. Kamihara {\it et al.}, Nature {\bf 453}, 376 (2008).

\bibitem{wen2} L. Shan {\it et al}, EPL {\bf 83}, 57004 (2008); G. Mu {\it et al}, Chin. Phys. Lett. {\bf 25}, 2221 (2008).

\bibitem{h1} X. H. Chen {\it et al.}, Nature {\bf 453}, 761 (2008).

\bibitem{h2} G. F. Chen {\it et al.}, 
Phys. Rev. Lett. {\bf 100}, 247002 (2008).

\bibitem{h3} Z. A. Ren {\it et al.},
Materials Research Innovations {\bf 12}, 105, (2008).

\bibitem{h4} Z. A. Ren {\it et al.},
Chin. Phys. Lett. {\bf 25}, 2215 (2008).

\bibitem{sdw1} J. Dong {\it et al.},
EPL {\bf 83}, 27006 (2008).

\bibitem{neutron1} Clarina de la Cruz {\it et al.},
  Nature {\bf 453}, 899 (2008).

\bibitem{neutron2} M. A. McGuire {\it et al.},
arXiv:0804.0796 (2008).

\bibitem{sing} D. J. Singh and M. H. Du, Phys. Rev. Lett. {\bf 100}, 237003 (2008).

\bibitem{xu} G. Xu {\it et al.},
EPL {\bf 82}, 67002 (2008).

\bibitem{mazin} I. I. Mazin, D.J. Singh, M.D. Johannes, and M.H.
Du, Phys. Rev. Lett. {\bf 101}, 057003 (2008).

\bibitem{kuroki} K. Kuroki {\it et al.}, 
Phys. Rev. Lett. {\bf 101}, 087004 (2008).

\bibitem{cao} C. Cao, P. J. Hirschfeld, and H. P. Cheng, Phys. Rev. B {\bf 77}, 220506(R) (2008).

\bibitem{dai} X. Dai, Z. Fang, Y. Zhou and F. C. Zhang, Phys. Rev. Lett. {\bf 101}, 057008 (2008).

\bibitem{han} Q. Han, Y. Chen, and Z. D. Wang, EPL {\bf 82}, 37007 (2008); arXiv:0803.4346 (2008).


\bibitem{Eremin}  M. M. Korshunov and I. Eremin, Phys. Rev. B {\bf 78}, 140509(R) (2008).

\bibitem{raghu} S. Raghu, X. L. Qi, C. X. Liu, D. J. Scalapino,
and S. C. Zhang, Phys. Rev. B {\bf 77}, 220503(R) (2008).


\bibitem{wxg} P. A. Lee and X. G. Wen, Phys. Rev. B {\bf 78}, 144517 (2008).

\bibitem{weng} Z. Y. Weng, arXiv:0804.3228 (2008).

\bibitem{ma} F. Ma, Z.Y. Lu, and T. Xiang, arXiv:0804.3370 (2008).

\bibitem{yin} Z. P. Yin {\it et al.}, Phys. Rev. Lett. {\bf 101}, 047001 (2008).

\bibitem{bickers} 
N. E. Bickers and D. J. Scalapino, Ann. Phys. (N.Y.) {\bf 193},
206 (1989); Z.-J. Yao, J.-X. Li, and Z. D. Wang, Phys. Rev. B {\bf
76}, 212506 (2007).

\bibitem{graser} S. Graser {\it et al.}, 
Phys. Rev. B {\bf 77}, 180514(R) (2008).

\bibitem{note} The effective  interaction strengths $(U, \, U^\prime, \, J, \, J^\prime) $  used here are not
equal to those defined in the orbital representation.

\bibitem{takimoto} T. Takimoto, T. Hotta, and K. Ueda, Phys. Rev. B
  {\bf 69}, 104504 (2004).


\bibitem{note2} For example, when $J^\prime =0.5$, the gap magnitudes
  of both bands decrease.

\bibitem{fawang} Fa Wang {\it et al.}, arXiv:0807.0498 (2008).

\bibitem{emerin1} We note that due to the essentially similar hole and
  electron Fermi pockets in the present model and that in Ref. 19,
  similar interband spin fluctuation
  peaks are observed, which was also thought to be relavent
  to the occurrence of the extended $s$-wave pairing in Ref. 19.



\bibitem{boeri} L. Boeri {\it et al.}, Phys. Rev. Lett. {\bf 101}, 026403 (2008).
\bibitem{ding} H. Ding {\it et al.}, EPL {\bf 83}, 47001 (2008).
\bibitem{qi} X.-L. Qi {\it et al.}, arXiv:0804.4332 (2008).



\end{thebibliography}
\end{document}